\documentclass[preprint,preprintnumbers,amssymb,prd,nofootinbib,tightenlines]{revtex4}
\usepackage{graphicx}
\usepackage{bm}

\newcommand{\ra}{\rightarrow}
\newcommand{\non}{\nonumber}

\newcommand{\cerr}[3]   {\mbox{${{#1}^{+ #2}_{- #3}}$}}
\newcommand{\aerrsy}[5] {\mbox{${{#1}^{+ #2 + #4}_{- #3 - #5}}$}}

\newcommand{\err}[3]   {\mbox{${{#1}\pm{#2}\pm{#3}}$}}
\def\etapr{{\eta^{\prime}}}

\def\nodata{}
\begin{document}
     
\baselineskip=15pt
\parskip=5pt
   
\preprint{UK/TP-2005-02} 

\vspace*{0.7in}  

\title{Towards a Precision Determination  
of $\alpha$ in $B\to\pi\pi$ Decays}
   
\author{Susan Gardner${}^{1}$}   
\email{gardner@pa.uky.edu}
   
\affiliation{
${}^1$Department of Physics and Astronomy, University of Kentucky, 
Lexington, Kentucky 40506-0055
\vspace{1ex} \\
}


\begin{abstract}
An assumption of isospin symmetry permits the determination
of $\sin(2\alpha)$ from the experimental study of $B\to\pi\pi$ decays. 
Isospin, however, is merely an approximate symmetry; its breaking
predicates a theoretical systematical error $\sigma_\alpha^{\rm IB}$
in the extraction of $\alpha$. 
We focus on the impact of $\pi^0-\eta,\eta^\prime$ mixing, 
as well as the manner in which 
it is amenable to empirical constraint, and determine
that $\sigma_\alpha^{\rm IB}$ can potentially be controlled to 
${\cal O}(1^\circ)$. 
\end{abstract}


\maketitle

\section{Introduction}   
Probing the mechanism of CP violation
in the $B$-meson system demands that the angles of the
unitarity triangle, $\alpha, \beta$, and $\gamma$, be extracted
a plurality of ways~\cite{review}. 
The study of $B\to \pi\pi$ decays, e.g., 
permits the determination of $\sin(2\alpha)$, where 
$\alpha\equiv \phi_2 \equiv \hbox{Arg}(- V_{tb}^* V_{td} / V_{ub}^* V_{ud})$
and we recall that $\alpha + \beta + \gamma=\pi\, ({\rm mod}\, 2\pi)$
in the Standard Model~\cite{sign}. 
The measurement of the time-dependent, CP-violating asymmetries 
in $B, \bar B \to \pi^+ \pi^-$ decays determine $S_{\pi\pi}$
and $C_{\pi\pi}$; these measurements in themselves determine 
$\sin(2\alpha_{\rm eff})$, where 
\begin{equation}
\sin(2\alpha_{\rm eff}) = \frac{S_{\pi\pi}}{\sqrt{1 - C_{\pi\pi}^2}} \,.
\end{equation}
Penguin contributions make the parameter $\alpha_{\rm eff}$ differ
from $\alpha$. Gronau and London have noted, however, that the 
``pollution'' $\Delta \alpha \equiv \alpha_{\rm eff} - \alpha$
can be determined
and removed with additional $B, \bar B\to \pi\pi$ data under 
an assumption of isospin symmetry~\cite{GL}. 
Isospin is an approximate symmetry --- the $u$ and $d$ quarks differ
in both their charge and mass; such isospin-breaking effects
make the determination of $\Delta \alpha$ imperfect. 
It is our purpose to study these effects, to the end of 
assessing 
the irreducible theoretical error $\sigma_{\alpha}^{\rm IB}$ 
in the determination of $\Delta \alpha$ 
via this method. Such is crucial to 
precision tests of the Standard Model of CP violation, realized 
through improved measurements at the current $B$-meson factories and
beyond~\cite{superB}. 
Earlier work has focused on 
the impact of $|\Delta I|=3/2$ 
electroweak penguins~\cite{ewpeng} 
and on the role of $\pi^0 -\eta, \eta^\prime$
mixing on $\Delta \alpha$~\cite{svgpieta}. 
The treatment of the latter, due to Gardner~\cite{svgpieta}, 
has recently
been reexamined by Gronau and Zupan~\cite{GZ}. 
The purpose of this article is to 
correct that work~\cite{GZ} 
and to update the analysis of Ref.~\cite{svgpieta}. 

\section{Isospin Analysis in $B\to\pi\pi$ Decay}
We begin by reviewing the isospin analysis in the 
isospin-perfect limit~\cite{GL}. In this limit, 
the $\pi^+, \pi^0$, and $\pi^-$ mesons form 
a degenerate isospin triplet, and the $B\to \pi^+ \pi^-$ decay amplitude 
must be symmetric under the exchange of the two pions, as per 
the constraint of Bose symmetry. We can relate the two-pion  
states to states of definite isospin $I$, $|(\pi\pi)_I\rangle$, via 
\begin{eqnarray}
| \pi^+\pi^-\rangle &\propto& 
|(\pi\pi)_0 \rangle 
+ \frac{1}{\sqrt{2}}  |(\pi\pi)_2 \rangle 
\non \\
| \pi^0\pi^0\rangle 
&\propto& 
|(\pi\pi)_0 \rangle 
- {\sqrt{2}}  |(\pi\pi)_2 \rangle 
\;, 
\end{eqnarray}
where the properly symmetrized state is 
$| \pi^-\pi^+ \rangle_{\rm sym} \equiv 
(| \pi_1^+\pi_2^- \rangle +  | \pi_1^-\pi_2^+ \rangle)/\sqrt{2}
= \sqrt{2} |\pi^+\pi^-\rangle$~\cite{svggv52}. The particles of the 
$\pi^0 \pi^0$ final state are identical, so that this state
need not be symmetrized -- if symmetrization is performed nevertheless, an
additional factor of $1/2$ must be applied to yield the correct
branching ratio. The $B^-, \bar B^0$ form 
a degenerate isospin doublet as well; 
evaluating the Clebsch-Gordon coefficients allows us to write a 
decomposition in terms of amplitudes $A_I$ of definite 
isospin~\cite{svggv52}:
\begin{eqnarray}
A_{B^0 \ra \pi^+\pi^-}\equiv
\langle \pi^+ \pi^- | {\cal H}_W | B^0\rangle
&\equiv& 
A_0 + \frac{1}{\sqrt{2}}  A_2 \;,
\non \\
A_{B^0 \ra \pi^0\pi^0}\equiv
\langle \pi^0 \pi^0 | {\cal H}_W | B^0\rangle
&\equiv& 
 A_0 
- {\sqrt{2}}  A_2 \;, 
\label{isodecomp}\\
A_{B^+ \ra \pi^+\pi^0}\equiv
\langle \pi^+ \pi^0 | {\cal H}_W | B^+\rangle
&\equiv&
\frac{3}{2} A_2   \;, 
\non
\end{eqnarray}
where analogous relationships in the charge-conjugate modes ($\bar A$) 
are implied. 
We recall that $A_0$ and $A_2$ 
are generated by 
$|\Delta I|=1/2$ and $|\Delta I|=3/2$ weak transitions, respectively. 
Since the symmetrized states 
$| \pi^-\pi^+ \rangle_{\rm sym}$ and $| \pi^+\pi^0 \rangle_{\rm sym}$
appear in the physical amplitudes, 
the 
$B\ra\pi^+\pi^-$ and $B\ra \pi^+ \pi^0$ partial widths are 
a factor of 2 larger than suggested by Eq.~(\ref{isodecomp})~\cite{branco}. 
We note that 
the reduced transition rate $\gamma(B\to \pi_1 \pi_2)$ is 
related to the partial width $\Gamma(B\to \pi_1 \pi_2)$ via 
\begin{equation}
\Gamma(B\to \pi_1 \pi_2) \equiv \frac{1}{16\pi M_B}
\sqrt{\left(1 - \frac{(M_{\pi_1} + M_{\pi_2})^2}{M_B^2}\right)
\left(1 - \frac{(M_{\pi_1} - 
M_{\pi_2})^2}{M_B^2}\right)} 
\gamma(B\to \pi_1 \pi_2) \,.
\label{defgamma}
\end{equation} 
Employing experimental masses throughout, we find, in specific, 
\begin{eqnarray}
\gamma_{+-} &\equiv& \gamma({B\to \pi^+\pi^-})=2|A_{B \ra \pi^+\pi^-}|^2 \,, 
\nonumber\\
\gamma_{+0} &\equiv& \gamma({B^+\to \pi^+\pi^0})=
2|A_{B^+ \ra \pi^+\pi^0}|^2 \,, \label{ampconv}\\
\gamma_{00} &\equiv& \gamma({B\to \pi^0\pi^0})=|A_{B \ra \pi^0\pi^0}|^2 \,.
\nonumber 
\end{eqnarray}
We note that electromagnetic radiative corrections, 
which may well be important, should be applied 
to yield the empirical decay widths and ultimately
the reduced transition rates~\cite{barachi}. 
Irrespective of such corrections, if we rewrite the amplitudes of 
Eq.~(\ref{isodecomp}), which satisfy the ``triangle relation,''
\begin{equation}
\frac{1}{\sqrt{2}}(A_{B\to \pi^+ \pi^-} - A_{B\to \pi^0 \pi^0}) 
= A_{B^+ \to \pi^+ \pi^0} \,,
\label{tri}
\end{equation} 
in terms of the amplitudes 
\begin{eqnarray}
A_{+-} &\equiv& \sqrt{2} A_{B\to \pi^+\pi^-} \,,
\nonumber\\
A_{00} &\equiv& - A_{B\to \pi^0\pi^0} \,,
\label{gzdef}\\
A_{+0} &\equiv& \sqrt{2} A_{B^+\to \pi^+\pi^0}
\,,
\nonumber 
\end{eqnarray}
defined so that $|A_{ij}|=\sqrt{\gamma_{ij}}$, 
we find 
\begin{equation}
A_{+-} +  \sqrt{2}A_{00} = \sqrt{2}A_{+0}\,. 
\label{trigamma}
\end{equation} 
We assume $\sqrt{2}|A_2| \ge |A_0|$, so that 
$A_{B\to \pi^0 \pi^0} \le 0$. This assumption is 
consistent with theoretical assays of $B\to \pi\pi$ decay
in an operator product expansion framework~\cite{th_rev},
as well as with the pattern of empirical branching ratios, given
current errors~\cite{hfag}.
Note that a similar triangle relation holds for the 
charge conjugate modes and thus 
\begin{equation}
\bar A_{+-} +  \sqrt{2}\bar A_{00} = \sqrt{2}\bar A_{-0}\,,
\label{trigammabar}
\end{equation} 
where we note $|\bar A_{-0}|^2 \equiv \gamma(B^- \to \pi^- \pi^0)$. 
The form of Eqs.~(\ref{trigamma}) and (\ref{trigammabar}) is identical
to that of earlier analyses~\cite{GL,svgpieta,GZ}, once differing
definitions are taken into account. 
The upshot of the 
isospin analysis is that 
the shift 
of $\alpha_{\rm eff}$ from $\alpha$ induced by the
penguin amplitude in $B\to\pi^+\pi^-$, namely 
\begin{equation}
\Delta \alpha \equiv \alpha_{\rm eff} - \alpha 
\equiv \frac{1}{2} \hbox{Arg}\left( e^{2i\gamma} \bar A_{+-}
A_{+-}^\ast \right) \,,
\end{equation}
can be expressed in terms of empirically determined quantities. 
In particular, with 
$\phi\equiv \hbox{Arg} (A_{+-} A_{+0}^\ast)$ and 
$\bar \phi\equiv \hbox{Arg} (\bar A_{+-} \bar A_{-0}^\ast)$, 
we have~\cite{GL,GZ}
\begin{equation}
\Delta \alpha_{\rm isospin} \equiv \frac{1}{2}(\bar \phi - \phi) 
= \frac{1}{2} \left( \hbox{Arg}(e^{2i\gamma} \bar A_{+-} A_{+-}^\ast)
- \hbox{Arg}(e^{2i\gamma} \bar A_{-0} A_{+0}^\ast)
\right) \,.
\label{dalpiso}
\end{equation}
In the isospin-perfect limit, as we examine here, 
$A_{+0}=\exp(2i\gamma)\bar A_{-0}$, so that the second term
vanishes and $\Delta \alpha_{\rm isospin} = \Delta \alpha$. 
Geometrically this implies that the two 
triangles share a common side; 
namely, $\sqrt{\gamma_{+0}} = \sqrt{\bar \gamma_{-0}}$. 
Let us now turn to an analysis of isospin-breaking effects. 
Interestingly, the most significant uncertainty arises from 
the manner in which 
$\Delta\alpha_{\rm isospin} \neq \Delta\alpha$, 
promoting
the importance of direct theoretical assays of 
$\Delta \alpha$~\cite{buch_safir}. 

\section{Isospin Breaking in $B\to\pi\pi$ Decay}
The charge and mass of the up and down quarks do differ, so 
that the predictions of the isospin analysis we have discussed
cannot strictly hold. There are two different effects to 
consider. Firstly, penguin contributions of $|\Delta I|=3/2$ character 
can occur, mediated either by electroweak penguin effects, or
by isospin-breaking in the strong-penguin 
matrix elements~\cite{burassil,svgpieta,svggv99}. 
Secondly, the triangle relationships of 
Eqs.~(\ref{trigamma}) and (\ref{trigammabar}) need no longer 
hold~\cite{svgpieta}. 
For example, the physical, neutral-pion state contains isoscalar
components due to mixing with the $\eta$ and $\eta^\prime$, engendering
an ``$I=1$'' amplitude in $B\to\pi\pi$ decay~\cite{svgpieta}. 
The $\eta$ and $\eta^\prime$
admixtures in the $\pi^0$ are generated by the strong interaction
in ${\cal O}(m_d-m_u)$.
Alternatively, we 
can regard this interaction as an $I=1$ ``spurion'', encoding
isospin-violating effects so that the matrix elements with the
spurion are $SU(2)_f$ invariant~\cite{tdlee}. In this latter 
picture, the ``extra'' amplitude 
engendered by $\pi^0 -\eta,\eta^\prime$ mixing can be recast as a
$|\Delta I|=5/2$ amplitude, generated by ${\cal O}(m_d - m_u)$ or
${\cal O}(\alpha)$ effects in concert with a $|\Delta I|=3/2$
weak transition. A $|\Delta I|=5/2$ transition can also be
realized by 
isospin-breaking effects in concert with 
a $|\Delta I|=1/2$ weak transition~\cite{branco}, though, 
as per the spurion picture, such 
is not engendered by 
$\pi^0-\eta,\eta^\prime$
mixing in leading order in isospin breaking. 
Writing $A_{|\Delta I|,I}$, we replace
Eq.~(\ref{isodecomp}) with~\cite{svggv52,svgulf} 
\begin{eqnarray}
A_{B^0 \ra \pi^+\pi^-}
&\equiv& 
A_{1/2,0}+ \frac{1}{\sqrt{2}}(A_{3/2,2}- A_{5/2,2}) \;,
\non \\
A_{B^0 \ra \pi^0\pi^0}
&\equiv& 
 A_{1/2,0} 
- {\sqrt{2}} ( A_{3/2,2} - A_{5/2,2}) \;, 
\label{isodecomp52}\\
A_{B^+ \ra \pi^+\pi^0}
&\equiv&
\frac{3}{2} A_{3/2,2}  + \sqrt{\frac{3}{2}} A_{5/2,2}  \;. 
\non
\end{eqnarray}
We note that this parametrization suffices to capture isospin
breaking in $B\to \pi\pi$ decay, as three theoretical amplitudes
describe the three empirical ones. 
Isospin breaking impacts the determination of $\Delta \alpha$ 
in two distinct ways. For example, it can break the triangle relation,
Eq.~(\ref{tri}). If the triangle relation is broken in an ill-determined
way, the ability to assess the angles $\phi$ and $\bar\phi$ is compromised. 
If the triangle relation is not broken, however, then the application 
of the isospin decomposition given in Eq.~(\ref{isodecomp}) permits
the determination of $\phi$ and $\bar\phi$ {\it regardless} of
whether additional isospin-breaking effects are present. 
Isospin breaking, however, can also make $\Delta \alpha_{\rm isospin}$ 
differ from $\Delta\alpha$; specifically, 
$\hbox{Arg}(e^{2i\gamma} \bar A_{-0} A_{+0}^\ast)\ne 0$,
recalling Eq.~(\ref{dalpiso}). If the impact of both effects can be estimated, 
if not controlled via empirical constraints, we can assess the irreducible
theoretical error in the determination of $\Delta\alpha$. 
Interpreting these two effects in terms of the parametrization of 
Eq.~(\ref{isodecomp52}), a non-zero value of the $A_{5/2,2}$ amplitude 
signals the breaking of the triangle relation, whereas 
penguin contributions to $A_{B^+ \to\pi^+\pi^0}$, to $A_{3/2,2}$, make 
$\hbox{Arg}(e^{2i\gamma} \bar A_{-0} A_{+0}^\ast)\ne 0$ even if
$A_{5/2,2}=0$. Electroweak penguin contributions are an example 
of the latter effect~\cite{kramer}. 
Since current experimental data is consistent with 
$|A_{3/2,2}|\gtrsim |A_{1/2,0}|$ in $B\to\pi\pi$ decay, we expect 
that $\pi^0-\eta,\eta^\prime$ mixing will play 
the most important role in the realization of a 
$A_{5/2,2}$ amplitude. 
In the limit that the $A_{5/2,2}$ amplitude is generated 
exclusively in this manner, 
$\hbox{Arg}(e^{2i\gamma} \bar A_{-0} A_{+0}^\ast)\ne 0$ can only be 
realized through penguin contributions of $|\Delta I|=3/2$ character. 
The phenomenon of $\pi^0-\eta,\eta^\prime$ mixing can generate both
effects; let us consider it explicitly. 

\subsection{$\pi^0-\eta,\eta^\prime$ Mixing}
In what follows we examine the
role of $\pi^0-\eta,\eta^\prime$ mixing on 
the extraction of $\alpha$ from $B\to\pi\pi$ decays. 
We distinguish the amplitude for decay to physical pion
final states, which suffer $\pi^0-\eta,\eta^\prime$ mixing, e.g., 
$A_{B\to \pi^0 \pi^0}$, from the amplitude in the isospin-perfect 
limit, $A_{B\to \phi_3 \phi_3}$, where $\phi_3$ denotes 
the isospin-triplet state with $I_3=0$. 
Noting earlier work on $\pi^0-\eta,\eta^\prime$ mixing
in $K\to\pi\pi$ decay~\cite{buras,trampetic,svggv52}, we have 
\begin{eqnarray}
A_{B^+ \to \pi^+ \pi^0} &=& A_{B^+ \to \pi^+ \phi_3} + 
\varepsilon A_{B^+\to \pi^+ \eta} + 
\varepsilon^\prime A_{B^+ \to \pi^+ \eta^\prime} \,, \nonumber \\
A_{B \to \pi^0 \pi^0} &=& A_{B \to \phi_3 \phi_3} + 
2\varepsilon A_{B\to \phi_3 \eta} + 
2\varepsilon^\prime A_{B\to \phi_3 \eta^\prime} \,,
\label{addpieta}
\end{eqnarray}
where we assert $\varepsilon\,,\varepsilon^\prime \sim {\cal O}((m_d-m_u)/\Lambda_{\rm had})$
or ${\cal O}(\alpha)$ and neglect all higher-order terms in
isospin-breaking parameters. To gain insight on the nature of
$\Lambda_{\rm had}$, we note that 
the analysis of the pseudoscalar
meson octet in current algebra~\cite{GTW}, or in lowest 
order chiral perturbation theory~\cite{GaL}, determine the
$\pi^0 - \eta_8$ mixing angle $\varepsilon_8$ to be 
\begin{equation}
\varepsilon_8 = \frac{\sqrt{3}}{4} \left(\frac{m_d - m_u}{m_s - \hat{m}}
\right) \,,
\end{equation}
with $\hat{m}=(m_u+ m_d)/2$, 
so that we expect $\Lambda_{\rm had}\sim {\cal O}(m_s)$. 
The impact 
of isospin breaking is controlled by the magnitude of
SU(3)$_f$ breaking. 
The breaking of SU(3)$_f$ symmetry also engenders
the mixing of the pseudoscalar octet and singlet states, 
$\eta_8$ and $\eta_0$, to realize the observed 
$\eta$ and $\eta^\prime$ states.
Such considerations 
demand that we evaluate 
$A_{B\to \pi \eta^{(\prime)}}$ in the presence of 
SU(3)$_f$ breaking effects. 
We postpone specific 
estimates of $\varepsilon$ and $\varepsilon^\prime$
to the discussion of our numerical results.  
We may use these relationships 
to rewrite the triangle relation, Eq.~(\ref{tri}), which now
appears as 
\begin{equation}
\frac{1}{\sqrt{2}}(A_{B\to \pi^+ \pi^-} - A_{B\to \phi_3 \phi_3}) 
= A_{B^+ \to \pi^+ \phi_3} \,,
\end{equation} 
in terms of amplitudes employing physical $\pi^0$ states. 
That is, 
\begin{eqnarray}
\frac{1}{\sqrt{2}}(A_{B\to \pi^+ \pi^-} - A_{B\to \pi^0 \pi^0}) 
&=& A_{B^+ \to \pi^+ \pi^0} - \sqrt{2} \varepsilon A_{B\to \phi_3\eta}
- \sqrt{2} \varepsilon^\prime A_{B\to \phi_3\eta^\prime}
\nonumber \\
&&-  \varepsilon A_{B^+\to \pi^+\eta}
-  \varepsilon^\prime A_{B^+\to \pi^+\eta^\prime}
\,,
\label{quad}
\end{eqnarray}
where replacing $A_{B\to \phi_3\eta^{(\prime)}}$ with
$A_{B\to \pi^0\eta^{(\prime)}}$ generates corrections
of higher 
order in $\varepsilon, \varepsilon^\prime$, which are negligible
in the order to which we work. 
Note that $\eta^{(\prime)}$
connotes either $\eta$ or $\eta^\prime$ throughout. 
We observe that the triangle relation is broken 
in the presence of isospin-breaking effects~\cite{svgpieta}. 

We turn to theory to assess the impact of the amplitudes
containing $\eta,\eta^\prime$ on Eq.~(\ref{quad}). 
The QCD factorization approach~\cite{BBNS} to hadronic $B$-meson decay 
analyzes the decay amplitudes 
in a systematic expansion in 
inverse powers of the heavy-quark mass $m_b$ 
and the strong coupling 
constant $\alpha_s(\mu)$, where $\mu \sim{\cal O}(m_b)$. 
Crucial to the treatment of the decay amplitudes in this case is 
that of the physical $\eta$ and $\eta^\prime$ states themselves,
as the $\eta$ and $\eta^\prime$ mix. 
These states are not simple flavor-octet and flavor-singlet states,
as SU(3)$_f$ symmetry would suggest, but rather each physical state
is a mixture of these components. The presence of the flavor-singlet
component in  
decays to $\eta^{(\prime)}$ final states admits novel decay mechanisms,
in part mediated
by the axial anomaly, 
not present in other channels~\cite{BNeta}.
We emphasize, as recognized in Ref.~\cite{BNeta},
that these flavor-singlet contributions are not captured by the analysis of
non-$\eta^{(\prime)}$ decay channels with an assumption of
SU(3)$_f$ symmetry. To implement $\eta-\eta^\prime$ mixing, we employ
the Feldmann-Kroll-Stech scheme~\cite{FKS}, also adopted in 
Refs.~\cite{BNeta,BN}, in which a single mixing
angle characterizes the decomposition of $|\eta^{(\prime)}\rangle$ 
into the flavor states 
$|\eta_q \rangle = (|u\bar u\rangle + |d\bar d\rangle)/\sqrt{2}$
and $|\eta_s \rangle = |s\bar s\rangle$. 
Beneke and Neubert
thus determine~\cite{BN} 
\begin{eqnarray}
&&\sqrt{2} A_{B^-\to\pi^-\eta^{(\prime)}}^{\rm BN} + 
2 A_{\bar B^0 \to \phi_3 \eta^{(\prime)}}^{\rm BN} 
= \sum_{p=u,c} \lambda_p \Bigg\{ A_{\pi \eta_q^{(\prime)}} 
\left[ \delta_{pu} (\beta_1 + \beta_2 + 2\beta_{S1} 
+ 2 \beta_{S2}) \right] \nonumber \\
&&+ \sqrt{2} A_{\pi\eta_s^{(\prime)}} 
\left[ \delta_{pu} (\beta_{S1} + \beta_{S2})
+ \frac{3}{2} \beta_{S3,EW}^p +  \frac{3}{2} \beta_{S4,EW}^p \right] 
\nonumber \\
&&+ A_{\eta_q^{(\prime)}\pi} 
\left[ \delta_{pu} (\alpha_1 + \alpha_2 + \beta_1 + \beta_2)
+ \frac{3}{2} \alpha_{3,EW}^p +  \frac{3}{2} \alpha_{4,EW}^p 
+ \frac{3}{2} \beta_{3,EW}^p +  \frac{3}{2} \beta_{4,EW}^p \right]\Bigg\} \,,
\label{sumamp}
\end{eqnarray}
where $\lambda_p\equiv V_{pb} V_{pd}^\ast$, and we refer to Ref.~\cite{BN} 
for all details. 
In this expression, the meson masses, decay constants, 
form factors, and light-cone distribution functions 
are all evaluated 
in the isospin-symmetric limit. Note, however, that 
the physical quark charges have
been employed, so that, in particular, $e_u \ne e_d$, 
in the evaluation of the electroweak penguin
contributions. We note that 
the role of a possible $c\bar c$ component in the 
$\eta^{(\prime)}$ mesons has been included in the computation
of $A_{B^-\to \pi^- \eta^{(\prime)}}$ and $A_{\bar B^0\to \phi_3 
\eta^{(\prime)}}$, 
although such effects are likely 
most significant in $b\to sq\bar q$ transitions~\cite{sjbsvg}, which
we do not treat here. 
The amplitudes are computed in next-to-leading order in $\alpha_s$
and at leading power in $\Lambda_{\rm QCD}/m_b$. 
The terms 
representing weak annihilation contributions 
are denoted by ``$\beta$'' and are 
included, although they are 
formally suppressed by a power of $m_b$. The computation of these
contributions suffer endpoint divergences in QCD factorization,
so that their estimate is uncertain. 
The large direct CP asymmetry found in the penguin-dominated mode
$B\to K^+\pi^-$~\cite{hfag} suggests that 
annihilation contributions may well play a larger phenomenological
role than anticipated~\cite{BN}. 
Nevertheless, if we do neglect
the annihilation contributions, as they are power-suppressed and 
largely possess, in this case, 
the same weak phase as the dominant contributions, 
we have 
\begin{equation}
\sqrt{2} A_{B^-\to\pi^-\eta^{(\prime)}}^{\rm BN} + 2 A_{\bar B^0 \to 
\phi_3 \eta^{(\prime)}}^{\rm BN} 
=\sum_{p=u,c} \lambda_p\left\{ A_{\eta_q^{(\prime)} \pi}
\left[ \delta_{pu} (\alpha_1 + \alpha_2) 
+ \frac{3}{2} \alpha_{3,EW}^p +  
\frac{3}{2} \alpha_{4,EW}^p \right]\right\} \,,
\label{sumeta}
\end{equation} 
where the $\alpha_i$ implicitly depend on the order of the
arguments of the $A_{ij}$ prefactor, so that 
$\alpha_i^{(p)}\equiv \alpha_i^{(p)}(\eta_q^{(\prime)} \pi)$. 
By comparison, we note that 
\begin{equation}
\sqrt{2} A_{B^-\to\pi^-\phi_3}^{\rm BN} = 
\sum_{p=u,c} \lambda_p \left\{A_{\pi\pi} 
\left[ \delta_{pu} (\alpha_1 + \alpha_2) 
+ \frac{3}{2} \alpha_{3,EW}^p +  \frac{3}{2} \alpha_{4,EW}^p \right]\right\}\,,
\label{ampBNpip0}
\end{equation}
where we emphasize 
that $\alpha_i^{(p)}\equiv \alpha_i^{(p)}(\pi \pi)$ in this case. 
The electroweak penguin contributions which appear in this expression
are explicitly of $|\Delta I|=3/2$ character. That is, 
were the quark-charge dependence made manifest, we would see 
that these contributions are proportional to $e_u - e_d$, so that,
in analogy to our discussion of $\pi^0-\eta,\eta^\prime$ mixing, 
the electroweak contribution contains an effective isovector
interaction acting in concert with a $|\Delta I|=1/2$ transition. 
Thus we can write
\begin{equation}
A_{B^- \to \pi^- \eta^{(\prime)}}^{\rm BN} 
+ \sqrt{2} A_{\bar B^0 \to \phi_3 \eta^{(\prime)}}^{\rm BN} 
=
A_{B^- \to\pi^- \phi_3}^{\rm BN} \bar X_{\eta_q^{(\prime)}} \;, 
\end{equation}
where 
\begin{equation}
\bar X_{\eta_q^{(\prime)}} = 
\left[\frac{A_{\eta_q^{(\prime)}\pi}}{A_{\pi\pi}}\right] 
\left[\frac{\bar \Sigma(\eta_q^{(\prime)} \pi)}{\bar \Sigma(\pi \pi)}
\right]
\label{Xdef}
\end{equation}
and 
\begin{equation}
\bar \Sigma(M_1 M_2) = \sum_{p=u,c} \lambda_p 
\left[\delta_{pu} (\alpha_1 + \alpha_2) 
+ \frac{3}{2} \alpha_{3,EW}^p +  \frac{3}{2} \alpha_{4,EW}^p\right]  \,,
\end{equation}
with $\alpha_i^{(p)}\equiv \alpha_i^{(p)}(M_1 M_2)$. 
We note $A_{B^i\to \pi^j \pi^k} = A_{B^i\to \pi^j \pi^k}^{\rm BN}/\sqrt{2}$, so
that the amplitudes 
$A_{B^i\to \pi^j \pi^k}^{\rm BN}$ satisfy both Eqs.~(\ref{isodecomp}) and 
(\ref{tri}), though the $A_I$ thus determined would be $\sqrt{2}$ larger.
Nevertheless, no physics can depend on this normalization choice, 
so that we must have 
\begin{equation}
\frac{|A_{B^-\to \pi^- \eta^{(\prime)}}^{BN}|}{|A_{B^-\to \pi^- \phi_3}^{BN}|}
= \frac{|A_{B^-\to \pi^- \eta^{(\prime)}}|}{|A_{B^-\to \pi^- \phi_3}|} 
\quad, \quad
\frac{|A_{\bar B^0\to \phi_3 \eta^{(\prime)}}^{BN}|}{|A_{\bar B^0\to \phi_3 \phi_3}^{BN}|}
= \frac{|A_{\bar B^0\to \phi_3 \eta^{(\prime)}}|}{|A_{\bar B^0\to \phi_3 \phi_3}|} \,,
\end{equation} 
as well as 
\begin{equation}
A_{B^- \to \pi^- \eta^{(\prime)}}
+ \sqrt{2} A_{\bar B^0 \to \phi_3 \eta^{(\prime)}}
=
A_{B^- \to\pi^- \phi_3} \bar X_{\eta_q^{(\prime)}} \;. 
\label{etatri}
\end{equation}
Returning to Eq.~(\ref{quad}), we find 
\begin{equation}
\frac{1}{\sqrt{2}}(A_{B\to \pi^+ \pi^-} - A_{B\to \pi^0 \pi^0}) 
= (1 - \xi )A_{B^+ \to \pi^+ \pi^0} \,,
\label{fixedtri}
\end{equation}
where $\xi$, which need not be real, is given by 
\begin{equation}
\xi =\varepsilon \left[ {X}_{\eta_q} + \dots \right]
+ \varepsilon^\prime \left[ {X}_{\eta_q^{\prime}} + \dots \right]
\,.
\label{xidef}
\end{equation}
We note that $\bar X_{\eta_q^{(\prime)}} \,\stackrel{CP}{\longleftrightarrow}\,
X_{\eta_q^{(\prime)}}$, 
$\bar \Sigma (M_1 M_2) \,\stackrel{CP}{\longleftrightarrow}\,
\Sigma (M_1 M_2)$, and 
 $\lambda_p \,\stackrel{CP}{\longleftrightarrow}\, \lambda_p^\ast$ under
$CP$ transformation, 
whereas each ellipsis denotes
neglected annihilation corrections. 
It is worth emphasizing that this result differs from its
analogue in Ref.~\cite{GZ} in an important way. 
That is, 
we have not assumed SU(3)$_f$ symmetry in the construction
of $\xi$, whereas Ref.~\cite{GZ} neglects all SU(3)$_f$-breaking
effects save for $\eta-\eta^\prime$ mixing. 
In Ref.~\cite{GZ}, $\xi$ is replaced by the parameter $e_0$, 
namely 
\begin{equation}
e_0 = \sqrt{\frac{2}{3}} \varepsilon  
+ \sqrt{\frac{1}{3}} \varepsilon^\prime \;.
\label{gztria} 
\end{equation} 
Let us examine the ingredients of Eqs.~(\ref{xidef}) and (\ref{Xdef}). 
The ratios of $\bar \Sigma(M_1 M_2)$ differ from unity 
if SU(3)$_f$ is broken in the light-cone distribution
functions, through, specifically, the hard-spectator contributions
to $\alpha_i(M_1 M_2)$~\cite{BN}. 
The latter are real if 
the contribution of the twist-3 distribution amplitudes, which 
generate divergent, albeit formally power-suppressed, contributions,
are neglected. This implies
that the ratios of $\bar \Sigma(M_1 M_2)$ in Eq.~(\ref{Xdef})
are real, if all subleading corrections and electroweak
penguin effects are neglected. The latter do make 
the ratios of $\bar \Sigma(M_1 M_2)$ complex in leading order
in $1/m_b$; however, electroweak penguin effects 
enter $\xi$ 
in ${\cal O}(\alpha (m_d - m_u)/\Lambda_{\rm had})$, so that their 
inclusion is actually of higher order in isospin breaking. 
The remaining factors in Eq.~(\ref{Xdef}) are given by 
\begin{equation} 
\left[\frac{A_{\eta_q^{(\prime)}\pi}}{A_{\pi\pi}}\right] = 
\frac{F_0^{B\to \eta^{(\prime)}}(0)}{F_0^{B\to \pi}(0)} \,,
\label{ffrat}
\end{equation} 
where $F_0^{B\to M}$ denotes a form factor for the decay
to a pseudoscalar meson $M$ in the convention of Bauer, Stech,
and Wirbel~\cite{BSW}. Following Beneke and Neubert, we 
parametrize~\cite{BNeta}
\begin{equation} 
F_0^{B\to \eta^{(\prime)}}(0) = 
F_1 
\frac{f_{\eta^{(\prime)}}^q}{f_\pi} 
+ F_2 \frac{\sqrt{2} f_{\eta^{(\prime)}}^q + f_{\eta^{(\prime)}}^s}
{\sqrt{3} f_\pi} \;,  
\label{ffetap}
\end{equation} 
noting that $F_1/F_2\sim {\cal O}(1)$ in the heavy-quark limit. 
The first term 
is related via SU(3)$_f$ breaking to 
$F_0^{B\to \pi}(0)$, where one expects $F_1 \approx F_0^{B\to \pi}(0)$ 
in the FKS scheme~\cite{BNeta}. The second term, however, 
is driven exclusively by the flavor-singlet contribution and cannot
be related to $F_0^{B\to \pi}(0)$. It is ill-known, though likely 
of greater impact on the $B\to \eta^\prime$ form factor~\cite{BNeta}. 

\subsubsection{Breaking the triangle relation}

A non-zero value of the parameter $\xi$ in Eq.~(\ref{fixedtri}) signals
the breaking of the triangle relation, Eq.~(\ref{tri}), and the appearance
of an amplitude of $|\Delta I|=5/2$ in character. In the QCD factorization
approach, $\xi$ is given by Eq.~(\ref{xidef}). If $\xi$ can be determined
with surety, and is real, an ``isospin analysis'' based on Eq.~(\ref{fixedtri})
can determine $\phi$ and $\bar \phi$ without theoretical error 
from this effect. 
Determining $\xi$ requires the 
isospin-breaking 
parameters
$\varepsilon$ and $\varepsilon^\prime$, which characterize
$\pi^0-\eta,\eta^\prime$ mixing, 
as well as $X_{\eta_q^{(\prime)}}$,
as per Eq.~(\ref{Xdef}). 
We begin by determining $X_{\eta_q^{(\prime)}}$ 
in the QCD factorization approach. 
The parameter $X_{\eta_q^{(\prime)}}$
is controlled by 
${F_0^{B\to \eta^{(\prime)}}}/{F_0^{B\to \pi}}$ 
and 
${\Sigma(\eta_q \pi)}/{\Sigma(\pi \pi)}$ exclusively, if 
power-suppressed contributions are indeed negligible. 
The former 
drives the numerical value of $\xi$. 
Using the 
parameters of Ref.~\cite{BNeta}, we find
\begin{eqnarray}
\frac{F_0^{B\to \eta}(0)}{F_0^{B\to \pi}(0)} &=& 
0.83 \pm 0.02 \to 0.89 
\;\; \left[ \sqrt{\frac{2}{3}} \approx 0.82\right]
\,,\nonumber \\
\frac{F_0^{B\to \eta^{\prime}}(0)}{F_0^{B\to \pi}(0)} 
&=& 0.68 \pm  0.02 \to 1.1 
\;\; \left[ \sqrt{\frac{1}{3}} \approx 0.58\right] 
\,,
\label{ffratres}
\end{eqnarray}
where the reported errors are determined from 
the errors in the inputs alone, assuming they are uncorrelated. 
The first number reported for each ratio employs $F_2=0$, whereas 
the second number employs, rather arbitrarily, $F_2=0.1$ as per 
Ref.~\cite{BNeta}.  
For reference, we have also included, in brackets, 
the ratios 
assumed in the SU(3)$_f$ approach of Ref.~\cite{GZ}. 
The form factor ratio for the $\eta^\prime$ can differ 
substantially from that 
of Ref.~\cite{GZ}, and varying 
the $\eta-\eta^\prime$ mixing angle does not
capture the excursion found. 
As for the remaining factor, 
we estimate, neglecting power corrections and
electroweak penguin contributions, 
\begin{eqnarray}
\frac{\Sigma(\eta_q \pi)}{\Sigma(\pi \pi)}
&\approx& 
\frac{\alpha_1(\eta_q^{(\prime)} \pi) + \alpha_2(\eta_q^{(\prime)} \pi)}
{\alpha_1(\pi \pi) + \alpha_2(\pi \pi)} \nonumber \\
&\approx& 1 + 
\frac{\alpha_s \pi f_{B_q}}{M_B \lambda_B}
\frac{f_\pi}{F_0^{B\to\pi}} (1 + \alpha_2^\pi)(\alpha_2^{\eta_q^\prime} 
- \alpha_2^\pi) \\ 
&\approx& 1 - 5\cdot 10^{-3} \nonumber \,,
\nonumber 
\end{eqnarray}
where the deviation from unity is determined by SU(3)$_f$
breaking 
in the light-cone distribution functions, as parametrized
by $\alpha_2^M$, which appear in the hard spectator
terms. We note, as in the case of annihilation contributions, 
that endpoint divergences can appear in the power
corrections. 
We employ the parameters given in Ref.~\cite{BN} and 
observe that this source of SU(3)$_f$ breaking appears to be
negligible. 
This observation is consistent with other recent data. For example, 
SU(3)$_f$ 
breaking in the
form factors and decay constants suffices to explain the
large difference in the observed
branching ratios for $B_s \to K^+ K^-$
and $B_d\to \pi^+ \pi^-$ decays~\cite{MB,pitts}.
We thus determine
\begin{eqnarray}
X_{\eta_q} 
&=& 
0.83 \pm 0.02 \to 0.89 
\,,\nonumber \\
X_{\eta_q^\prime} 
&=& 0.68 \pm 0.02  \to 1.1 
\label{Xres}
\,. 
\end{eqnarray} 
It is worth noting that 
the form of Eq.~(\ref{etatri}) 
is quite general; it does not rely on our adopted framework 
for $\eta-\eta^\prime$ mixing. 
If, instead, a general, two-angle mixing formalism~\cite{twoangle,EF} 
in the octet-singlet basis 
were employed to describe $\eta-\eta^\prime$ mixing, 
an equation of form Eq.~(\ref{etatri}) 
would nevertheless emerge~\cite{Ali98}. We would also
find compatible numerical results~\footnote{
In a two-angle mixing formalism, assuming as per Ref.~\cite{BNeta},
that the ratios of the $B\to \eta^{(\prime)}$ 
and $B\to \pi$ 
form factors are determined
by the ratios of the related decay constants, we have 
\begin{eqnarray} 
\frac{F_0^{B\to \eta}(0)}{F_0^{B\to \pi}(0)} &=& 
\frac{\sqrt{2}}{f_\pi}\left( \frac{f_8 \cos \theta_8}{\sqrt{6}} 
- \frac{f_0 \sin \theta_0}{\sqrt{3}} 
\right)
\,,\nonumber \\
\frac{F_0^{B\to \eta^{\prime}}(0)}{F_0^{B\to \pi}(0)} 
&=& 
\frac{\sqrt{2}}{f_\pi}\left( \frac{f_8 \sin \theta_8}{\sqrt{6}} 
+ \frac{f_0 \cos \theta_0}{\sqrt{3}} \right)
\,,
\end{eqnarray}
If $f_8=f_0=f_\pi$ and $\theta_8=\theta_0=\theta$ and
we assume the ideal mixing angle $\theta=\sin^{-1}(-1/3)$
we recover 
$F_0^{B\to \eta}(0)/F_0^{B\to \pi}(0) = \sqrt{2/3}$ 
and $F_0^{B\to \eta^\prime}(0)/F_0^{B\to \pi}(0) = \sqrt{1/3}$ 
as per Ref.~\cite{GZ}. If we employ the parameters in either
Eq.(3.7) or Eq.(3.8) of Ref.~\cite{EF} 
we find results comparable to what we have reported
in the $F_2=0$ case.}.
Indeed, we can use the empirical decay amplitudes to 
define and determine 
an effective parameter $\bar X_{\eta_q^{(\prime)}}^{\rm eff}$
via
\begin{equation}
A_{B^- \to \pi^- \eta^{(\prime)}}
+ \sqrt{2} A_{\bar B \to \pi^0 \eta^{(\prime)}}
=
A_{B^- \to\pi^- \pi^0} \bar X_{\eta_q^{(\prime)}}^{\rm eff} \;, 
\label{empX}
\end{equation} 
where if power corrections, as well as isospin-breaking effects,
are negligible, 
$\bar X_{\eta_q^{(\prime)}}^{\rm eff}$ is $\bar X_{\eta_q^{(\prime)}}$ 
as defined in Eq.~(\ref{Xdef}). 
Empirical branching ratios
for  $B^-\to\pi^-\eta^{(\prime)}$, $\bar B\to\pi^0\eta^{(\prime)}$, 
and $B^- \to \pi^- \pi^0$ decays 
can thus eventually determine 
$|\bar X_{\eta_q^{(\prime)}}^{\rm eff}|$, and the angles 
$\hbox{Arg}(A_{B^-\to\pi^-\eta^{(\prime)}}A_{B^-\to\pi^-\pi^0}^\ast 
\bar X_{\eta_q^{(\prime)}}^{{\rm eff},\ast})$ and 
$\hbox{Arg}(A_{\bar B\to\pi^0\eta^{(\prime)}}A_{B^-\to\pi^-\pi^0}^\ast
\bar X_{\eta_q^{(\prime)}}^{{\rm eff},\ast})$, up to discrete ambiguities. 
Data on the charge 
conjugate modes would determine the charge-conjugates of these 
quantities in a similar manner. Our theoretical analysis suggests that 
$\bar X_{\eta_q^{(\prime)}}^{\rm eff}$ is real to a good 
approximation\footnote{The SU(3)$_f$-breaking electroweak
penguin contribution to $\Sigma(\eta_q^{(\prime)})/\Sigma(\pi\pi)$
generates the only complex contribution in leading power in $1/m_b$. 
Note that of the neglected annihilation terms, only the 
electroweak penguin annhilation contributions can be 
complex. 
}, 
so that the deduced empirical angles can be interpreted as 
$\hbox{Arg}(A_{B^-\to\pi^-\eta^{(\prime)}}A_{B^-\to\pi^-\pi^0}^\ast)$ 
and 
$\hbox{Arg}(A_{\bar B\to\pi^0\eta^{(\prime)}}
A_{B^-\to\pi^-\pi^0}^\ast)$~\cite{GZ}. Nevertheless, verifying that
$|\bar X_{\eta_q^{(\prime)}}^{\rm eff}|= |X_{\eta_q^{(\prime)}}^{\rm eff}|$,
e.g., 
would serve as a consistency check. 
We thus expect that this analysis 
would not only constrain the ill-known $B\to \eta^{(\prime)}$ form 
factors, but also help determine the extent to which 
$\Delta \alpha_{\rm isospin} \ne \Delta \alpha$, as we shall explain.

\begin{table}[t]   
\caption{\label{branch}%
CP-averaged branching ratios for selected 
$B\to PP$ modes from the compilation of 
Ref.~\cite{hfag}, reported as 
$10^6\,\, \hbox{Br}({B\to PP})$. We display 
the experimental data, both preliminary
and published, available since the compilation
of Ref.~\cite{pdg2004} and included in the
averages of Ref.~\cite{hfag}. 
} \centering   \footnotesize
\vskip 0.5\baselineskip
\begin{tabular}{@{\hspace{3ex}}c@{\hspace{3ex}}
c@{\hspace{3ex}}c@{\hspace{3ex}}c@{\hspace{3ex}}
c@{\hspace{3ex}}c@{\hspace{3ex}}c@{\hspace{3ex}}
}
\hline \hline  
$\begin{array}{c}\displaystyle \mbox{Mode}
 \\ \displaystyle PP \end{array}$  
&  
PDG~\cite{pdg2004} 
&  
BABAR
&  
Belle 
&
CLEO 
&
CDF~\cite{cdf}
&
HFAG~\cite{hfag}
\\ \hline & & & & & & \vspace{-3ex} \\
$\pi^+\pi^0$                                      & 
$\cerr{5.6}{0.9}{1.1}$                            & 
{$\err{5.8}{0.6}{0.4}$}~\cite{ba1}                      & 
{$\err{5.0}{1.2}{0.5}$}~\cite{be1}                       & 
$\aerrsy{4.6}{1.8}{1.6}{0.6}{0.7}$~\cite{cl1}                & 
\nodata                                           & 
$5.5 \pm 0.6$                                     \\

$\eta\pi^+$                                       & 
$<5.7$                                            & 
{$\err{5.1}{0.6}{0.3}$}~\cite{ba2}                      & 
{$\err{4.8}{0.7}{0.3}$}~\cite{be2}                      & 
\cerr{1.2}{2.8}{1.2}~\cite{cl2}                              & 
\nodata                                           & 
$4.9 \pm 0.5$                                     \\

$\etapr\pi^+$                                     & 
$<7$                                              & 
{$\err{4.0}{0.8}{0.4}$}~\cite{ba2}                      & 
$<7$~\cite{be3}                                              & 
$1.0^{+5.8}_{-1.0}$~\cite{cl2}                               & 
\nodata                                           & 
$4.0 \pm 0.9$                                     \\

$\pi^+\pi^-$                                      & 
$4.8\pm0.5$                                       & 
$\err{4.7}{0.6}{0.2}$~\cite{ba3}                  & 
{$\err{4.4}{0.6}{0.3}$}~\cite{be1}            & 
$\aerrsy{4.5}{1.4}{1.2}{0.5}{0.4}$~\cite{cl1}     & 
{$4.4\pm1.3$} 
                       & 
$4.5 \pm 0.4$                                     \\

$\pi^0\pi^0$                                      & 
$1.9\pm 0.5$                                      & 
{$\err{1.17}{0.32}{0.10}$}~\cite{ba1}        & 
{$\aerrsy{2.3}{0.4}{0.5}{0.2}{0.3}$}~\cite{be4}         & 
$<4.4$~\cite{cl1}                                            & 
\nodata                                           & 
$1.45 \pm 0.29$                                   \\

$\eta\pi^0$                                       & 
$<2.9$                                            & 
{$<2.5$}~\cite{ba4}                                      & 
{$<2.5$}~\cite{be2}                                     & 
$<2.9$~\cite{cl2}                                            & 
\nodata                                           & 
{$<2.5$}                                          \\

$\etapr\pi^0$                                     & 
$<5.7$                                            & 
{$<3.7$}~\cite{ba4}                                      & 
\nodata                                           & 
$<5.7$~\cite{cl2}                                            & 
\nodata                                           & 
{$<3.7$}                                          \\

\hline \hline  
\end{tabular}
\vspace{1ex} \\  
\end{table}

Before turning to this issue, let us conclude by determining 
the expected value of the $|\Delta I|=5/2$ parameter $\xi$ and
the manner in which its uncertainty impacts $\Delta\alpha_{\rm isospin}$. 
To compute $\xi$, we use the recent results of Kroll for 
the $\pi^0-\eta,\eta^\prime$ mixing angles~\cite{kroll}: 
\begin{equation} 
\varepsilon = 0.017 \pm 0.003 \quad \,; \quad
\varepsilon^\prime = 0.004 \pm 0.001 \,,
\label{epsnos}
\end{equation}
to yield 
\begin{equation} 
\xi = 0.017 \pm  0.003 \rightarrow 0.020 \quad [0.016] \,,
\end{equation}
where the error in $\xi$ is determined from those of 
 the inputs alone, assuming their errors are uncorrelated. 
We have incorporated the $\pi^0-\eta,\eta^\prime$ mixing
angles directly as determined in low-energy experiments; we
note that the scale dependence of the light-cone distribution
functions, of which this is part, does not enter at next-to-leading order
accuracy in $\alpha_s$~\cite{BBNS}. 
The results employ $F_2$ as in Eq.~(\ref{ffratres}) and
report, in brackets, the value found in the
SU(3)$_f$ analysis of Ref.~\cite{GZ} as well. Note
that the greater uncertainty in $X_{\eta_q^\prime}$ noted 
previously has little bearing on the final error in $\xi$,
as the $\pi^0-\eta^\prime$ mixing angle $\varepsilon^\prime$ is 
relatively small. 
Given an estimate of $\xi$ and its error, 
we can also proceed to determine the uncertainty 
in $\phi$ consequent to it. 
To do this, we note that $\cos\phi$ can be determined
from the empirical decay amplitudes, determined from the 
empirical branching ratios via Eqs.~(\ref{defgamma},\ref{ampconv}),
and the relationship given in Eq.~(\ref{fixedtri})\footnote{The sign of $\phi$ 
is undetermined.}. 
We employ Eq.~(\ref{isodecomp}) for the neutral modes, but
define in this case
\begin{equation}
|A_2| = \frac{2}{3}(1- \xi) |A_{B^+ \to \pi^+ \pi^0}| \,,
\end{equation}
to yield 
\begin{equation}
\cos\phi = \cos\phi_0 +  \xi 
\left(\cos\phi_0 - \sqrt{2} 
\frac{|A_{B^+\to\pi^+ \pi^0}|}{|A_{B\to\pi^+ \pi^-}|}
\right) + {\cal O}(\xi^2) \,,
\end{equation} 
where 
\begin{equation}
\cos\phi_0 =\frac{1}{2\sqrt{2}} 
\left[\frac{
|A_{B\to\pi^+ \pi^-}|^2 - |A_{B\to\pi^0 \pi^0}|^2 + 
2 |A_{B^+\to\pi^+ \pi^0}|^2}
{|A_{B^+\to \pi^+ \pi^0}||A_{B\to\pi^+ \pi^-}|}  
\right] \,.
\end{equation} 
The error in $\xi$ generates an error in the determination
of $\phi$; namely, 
$\sigma_\phi = \sigma_\xi |\partial \phi/\partial \xi|$,
where we note that the error in $\bar\phi$ follows
from replacing the amplitudes by their CP conjugates. 
Assuming a 100\% error in our estimate of $\sigma_\xi$, as
$F_2$ is ill-known and the $\pi^0-\eta,\eta^\prime$ mixing
angles can have a small electromagnetic component, estimated to 
be some 6\% of $\varepsilon_8$~\cite{trampetic}, 
we employ $\sigma_\xi=0.006$  and a recent empirical compilation
of CP-averaged branching ratios~\cite{hfag}, reported in 
Table \ref{branch}, 
to estimate $\sigma_\phi=0.4^\circ$ and thus an error in 
$\Delta\alpha_{\rm isospin}$ of $0.4^\circ$, as we add the
errors linearly. 
In constrast, the shift in $\phi$ due to the ${\cal O}(\xi)$
contribution is $1.2^\circ$. 
We note, in particular, that 
our error estimate can be made more robust, if not reduced, 
through the measurement of $B\to\pi \eta^{(\prime)}$ decays.

\vspace{0.75cm}
\begin{figure}[ht]
\includegraphics[width=3in]{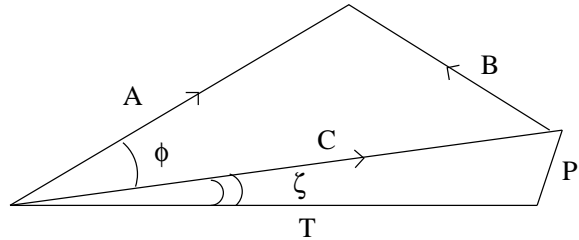}
\vspace{-0.2cm}
\caption{\label{figtri}
Schematic illustration of the triangle relation in $B\to\pi\pi$
decay in the presence of isospin breaking, Eq.~(\protect{\ref{fixedtri}}). 
Note that $A\equiv A_{B\to \pi^+\pi^-}/\sqrt{2}$, 
$B\equiv A_{B\to \pi^0\pi^0}/\sqrt{2}$, and 
$C\equiv (1- \xi)A_{B^+\to\pi^+ \pi^0}$. 
Moreover, $\xi$ is real, with $A_{B^+\to\pi^+ \pi^0}= T + P$, where 
$T$ is the tree-level contribution to $B^+\to\pi^+\pi^0$ decay
in the isospin-perfect limit. 
}
\end{figure}

\subsubsection{Breaking $\Delta\alpha_{\rm isospin} = \Delta\alpha$}

Thus far we have determined the error in $\Delta \alpha_{\rm isospin}$
incurred through the uncertainty in the parameter $\xi$. The error 
in $\Delta \alpha$, however, is determined by that in 
\begin{equation}
\Delta \alpha = 
\Delta \alpha_{\rm isospin} + \frac{1}{2}\hbox{Arg}(e^{2i\gamma} \bar A_{-0} A_{+0}^\ast)
\,.
\label{angrel}
\end{equation}
Defining
$\zeta\equiv \hbox{Arg} (A_{B^+\to\pi^+ \pi^0} T^\ast)$ and 
$\bar \zeta\equiv \hbox{Arg} (\bar A_{B^- \to \pi^- \pi^0} \bar T^\ast)$, 
where $T$ is the tree-level contribution to $B^+\to \pi^+ \pi^0$ 
decay in the isospin-perfect limit, so that $|T|=|\bar T|$, 
we have 
\begin{equation}
\hbox{Arg}(e^{2i\gamma} \bar A_{-0} A_{+0}^\ast) = 
\bar \zeta - \zeta \,.
\end{equation}
The interplay of these relationships is illustrated in 
Fig.~\ref{figtri}. As in the isospin-perfect case, the
angle $\phi$ has a discrete ambiguity, as does the angle $\zeta$:  
their overall sign is not determined. This 
is realized as an ambiguity in the orientation of 
$\triangle ABC$, that is, whether it 
points up or down. 
A similar ambiguity also exists for 
the charge-conjugate amplitudes, which yield $\bar\phi$
and $\bar \zeta$, to yield a four-fold ambiguity in 
$\Delta \alpha$. The angles $\zeta$ and $\bar \zeta$ are non-zero if 
penguin contributions of $|\Delta I|=3/2$ character occur. 
We wish to estimate the extent to which 
$\Delta \alpha \ne \Delta \alpha_{\rm isospin}$, as well as its uncertainty,  
though we shall begin by considering 
the contribution from $\pi^0-\eta,\eta^\prime$ mixing exclusively. 

In the presence of $\pi^0-\eta,\eta^\prime$ mixing, we have 
\begin{equation}
A_{B^+\to\pi^+ \pi^0} = A_{B^+\to\pi^+ \phi_3} 
+ \varepsilon A_{B^+\to\pi^+ \eta} 
+ \varepsilon^\prime A_{B^+\to\pi^+ \eta^\prime} \,, 
\end{equation}
with $A_{B^+\to\pi^+ \phi_3} = T + P_{\rm ew}$, 
where 
we emphasize that the amplitude computed 
with the $I=1$, $I_3=0$ state $\phi_3$
can contain an $|\Delta I|=3/2$ electroweak 
penguin contribution, $P_{\rm ew}$,
in addition to a tree-level contribution $T$. 
The angle $\zeta$ can be written as
\begin{equation}
\zeta=\hbox{Arg}(A_{B^+\to\pi^+\pi^0} A_{B^+\to\pi^+\phi_3}^\ast) 
+ \hbox{Arg}(A_{B^+\to\pi^+\phi_3} T^\ast) \,,
\label{defzeta}
\end{equation}
where the first term is of ${\cal O}(\varepsilon, \varepsilon^\prime)$ 
and the second
is rendered non-zero by $P_{\rm ew}$. Here we focus on the first term, namely, 
\begin{eqnarray}
\zeta_{\eta,\eta^\prime}&\equiv&
\hbox{Arg}(A_{B^+\to\pi^+\pi^0}A_{B^+\to\pi^+\phi_3}^\ast)
= \varepsilon \sin\theta_\eta 
\frac{|A_{B^+\to\pi^+\eta}|}{|A_{B^+\to\pi^+\pi^0}|}
+ \varepsilon^\prime \sin\theta_{\eta^\prime} 
\frac{|A_{B^+\to\pi^+\eta^\prime}|}{|A_{B^+\to\pi^+\pi^0}|} 
\,,\nonumber \\
&=& \varepsilon \sin\theta_\eta 
\sqrt{\frac{\gamma(B^+\to\pi^+\eta)}{\gamma(B^+\to\pi^+\pi^0)}} + 
\varepsilon^\prime \sin\theta_{\eta^\prime} 
\sqrt{\frac{\gamma(B^+\to\pi^+\eta^\prime)}{\gamma(B^+\to\pi^+\pi^0)}}  
\,,
\end{eqnarray}
where we work in ${\cal O}(\varepsilon,\varepsilon^\prime)$ throughout, with 
\begin{equation}
\theta_{\eta^{(\prime)}}\equiv 
\hbox{Arg}(A_{B^+\to\pi^+\eta^{(\prime)}}A_{B^+\to\pi^+\pi^0}^\ast) 
\,.
\end{equation}
Defining analogous variables for the CP-conjugate amplitudes, we determine 
that the contribution to $\Delta \alpha - \Delta \alpha_{\rm isospin}$ from
$\pi^0-\eta,\eta^\prime$ mixing is 
\begin{eqnarray}
\frac{1}{2}(\bar \zeta_{\eta,\eta^\prime} - \zeta_{\eta,\eta^\prime}) 
&=& \frac{\varepsilon}{2}
\left(
\sin\bar\theta_\eta 
\sqrt{\frac{\gamma(B^-\to\pi^-\eta)}{\gamma(B^-\to\pi^-\pi^0)}} 
- \sin\theta_\eta 
\sqrt{\frac{\gamma(B^+\to\pi^+\eta)}{\gamma(B^+\to\pi^+\pi^0)}} \right) \nonumber \\
&+& 
\frac{\varepsilon^\prime}{2}
\left( 
\sin\bar\theta_{\eta^\prime} 
\sqrt{\frac{\gamma(B^-\to\pi^-\eta^\prime)}{\gamma(B^-\to\pi^-\pi^0)}}  
- 
\sin\theta_{\eta^\prime} 
\sqrt{\frac{\gamma(B^+\to\pi^+\eta^\prime)}{\gamma(B^+\to\pi^+\pi^0)}} 
\right) \,.
\end{eqnarray}
Letting $\sin\bar\theta_{\eta^{(\prime)}}=-\sin\theta_{\eta^{(\prime)}}=1$ and
employing the empirical, 
CP-averaged branching ratios of the compilation of Ref.~\cite{hfag}, 
we estimate 
\begin{equation}
\frac{1}{2}(\bar \zeta_{\eta,\eta^\prime} - \zeta_{\eta,\eta^\prime}) 
\le 1.1^\circ \pm 0.2^\circ 
\,, 
\end{equation}
where the error follows from the errors in the branching ratios and
$\pi^0-\eta,\eta^\prime$ mixing angles alone, 
assuming
such are uncorrelated. Note that we have added the errors in 
$\bar \zeta_{\eta,\eta^\prime}$ and $\zeta_{\eta,\eta^\prime}$ 
linearly. 
This numerical result 
can be compared to 
the bound of $1.6^\circ$
at 90\% confidence level (CL) reported in Ref.~\cite{GZ}; 
our bound is slightly smaller as it employs measured branching
ratios, rather than experimental bounds. 
Its provenance is also different, as the bound
in our case does not depend on an assertion 
of SU(3)$_f$ symmetry. Most importantly, it is improvable, as
it is driven by empirical errors. In addition, 
the angles 
$\bar\theta_{\eta^{(\prime)}}$ and $\theta_{\eta^{(\prime)}}$
are also subject to empirical constraint, so that a direct 
assessment of $(\bar \zeta_{\eta,\eta^\prime} - \zeta_{\eta,\eta^\prime})/2$
should eventually prove possible. 
Ultimately, it is the 
uncertainty in $(\bar\zeta_{\eta,\eta^\prime} - \zeta_{\eta,\eta^\prime})/2$ 
which matters, not its gross deviation from zero.

\subsection{Other Isospin-Breaking Effects}

Thus far we have considered the disparate roles of $\pi^0-\eta,\eta^\prime$ 
mixing:
this isospin-breaking effect can not only engender 
a $|\Delta I|=5/2$ amplitude, 
breaking the triangle relation of Eq.~(\ref{tri}), 
but also generate a $|\Delta I|=3/2$ penguin amplitude, 
forcing 
$\Delta \alpha_{\rm isospin} - \Delta \alpha \ne 0$. Yet isospin breaking is 
not limited to $\pi^0-\eta,\eta^\prime$ mixing, and we can ask
what other effects might enter, as well as how we might discern their presence 
from the experimental data. 

As we have mentioned, $|\Delta I|=3/2$ 
electroweak penguin contributions can contribute
to $\Delta \alpha_{\rm isospin} - \Delta \alpha \ne 0$, through 
the second term of Eq.~(\ref{defzeta}). If one neglects the
electroweak penguin 
operators associated with small Wilson coefficients, namely $c_7$
and $c_8$~\cite{th_rev}, 
then the impact of these contributions on $\Delta \alpha$
can be assessed, without theoretical ambiguity, 
up to isospin-violating corrections~\cite{ewpeng},  
to yield~\cite{GZ}
\begin{equation}
\left( \Delta \alpha - \Delta \alpha_{\rm isospin} \right)_{\rm ewp} 
= 1.5^\circ \pm 0.3^\circ \,,
\end{equation} 
where the error arises from that in the empirical inputs. 
Isospin-breaking in the matrix elements of the
strong penguin operators can also engender a $|\Delta I|=3/2$ 
contribution~\cite{svgpieta,burassil}, not captured by 
$\pi^0-\eta,\eta^\prime$ mixing. 
For example, 
corrections of
${\cal O}(\alpha)$ can distinguish $A_{\pi^\pm\phi_3}$ from 
$A_{\phi_3\pi^\pm}$, or, specifically, 
$F_0^{B^\pm\to \pi^\pm}(0)f_{\phi_3}$
from $F_0^{B^\pm\to \phi_3}(0)f_{\pi^\pm}$. In addition, $m_d\ne m_u$ effects
beyond $\pi^0-\eta,\eta^\prime$ mixing can also 
occur~\cite{svggv99,burassil}, though in $K\to\pi\pi$ 
decay, e.g., such terms do not 
appear in the weak chiral Lagrangian in ${\cal O}(p^2)$~\cite{cronin}. 
The contributions from electroweak
penguin operators should yield the 
largest effect~\cite{svgpieta}. In particular, 
we note, using the notation of
Ref.~\cite{BBNS}, that $\alpha |C_4| /|C_9| \sim 4\%$.

Contributions to the $|\Delta I|=5/2$ amplitude, not 
mediated by $\pi^0-\eta,\eta^\prime$ mixing, 
can also occur. 
For example, ${\cal O}(\alpha)$ 
effects in the evaluation of $F_0^{B\to\pi}$ and $f_\pi$ can yield
an effective $|\Delta I|=5/2$ amplitude from either 
$|\Delta I|=3/2$ or $|\Delta I|=1/2$ 
weak transition operators. 
Contributions built on the former can be absorbed by 
modifying $\xi$ and $\bar \xi$ and enlarging their
errors. 
Contributions built on the latter
are more problematic, as they 
will make $\xi$ and $\bar \xi$ complex, as well as $\xi\ne\bar\xi$. 
Consequently, the angle determined from the analysis of 
Eq.~(\ref{fixedtri}), i.e., 
$\phi^\prime\equiv \hbox{Arg}(A_{+-}(1-\xi^\ast) 
A_{B^+\to \pi^+\pi^0}^\ast)$,
is not $\phi$. A similar conclusion emerges from the
study of the charge conjugate amplitudes, where we note 
$\bar\phi^\prime\equiv \hbox{Arg}(\bar A_{+-}(1- \bar\xi^\ast)
A_{B^+\to \pi^+\pi^0}^\ast)$ is not $\bar\phi$. Generally we can rewrite
Eq.~(\ref{angrel}) 
as 
\begin{equation}
\Delta \alpha = \frac{1}{2}(\bar \phi^\prime - \phi^\prime) 
+ \frac{1}{2}(\bar \zeta - \zeta)  
+ \frac{1}{2}\left[(\bar \phi - \phi) - (\bar \phi^\prime - \phi^\prime)\right]
\,,
\end{equation}
where the last term vanishes if $\xi$ and $\bar\xi$ are real. 
Note that the geometric interpretation of $\phi$ and $\zeta$ illustrated
in Fig.~\ref{figtri}, as well as of $\bar \phi$ and $\bar\zeta$, make
the signed contributions of 
$(\bar \phi - \phi)/2$ and $(\bar \zeta - \zeta)/2$, 
and, by
inference, $(\bar \phi^\prime - \phi^\prime)/2$ and $(\bar \zeta - \zeta)/2$, 
 add constructively. 
The sign of the term 
$[(\bar \phi - \phi) - (\bar \phi^\prime - \phi^\prime)]/2$, however,
is unclear. Nevertheless, 
in contradistinction to $K\to\pi\pi$ decays~\cite{svggv52,svgem,ciri}, 
we do expect the role of the $|\Delta I|=1/2$ 
weak transition in generating an effective $|\Delta I|=5/2$ amplitude 
in $B\to\pi\pi$ decays to be a relatively small effect. 
That is, on general grounds, the pattern of empirical branching ratios shows
that the $|\Delta I|=1/2$ amplitude is not dominant, indeed
that $|A_{3/2,2}|\gtrsim |A_{1/2,2}|$, and  
$\alpha/\varepsilon \approx 0.4$. 
It is worth noting, though, 
that $\xi$ and $\bar \xi$ can be complex from the 
inclusion of $\pi^0-\eta,\eta^\prime$ effects alone. However,
such effects arise, in leading power, from SU(3)$_f$ breaking 
in the light-cone distributions functions of the $\eta^{(\prime)}$
and $\pi$ which appear in the electroweak penguin contributions 
and, in the power corrections, through 
the electroweak penguin annihilation contributions. 
We find the leading-power effect to be negligibly small, though
testing, as per Eq.~(\ref{empX}), whether
$|X_{\eta_q^{(\prime)}}^{\rm eff}|=|\bar X_{\eta_q^{(\prime)}}^{\rm eff}|$
is bourne out by experiment should reveal the presence of
unexpectedly large complex contributions.

\section{Summary}

The study of $B\to\pi\pi$ decays under an assumption of isospin
symmetry permits the extraction of the angle 
$\alpha$, modulo discrete ambiguities. 
This is realized through the 
determination of 
the penguin pollution
$\Delta \alpha$, which is also discretely ambiguous, 
yielding $\alpha = \alpha_{\rm eff} - \Delta \alpha$ 
from the directly measured quantity $\sin (2\alpha_{\rm eff})$. 
Isospin symmetry is broken in nature, as the up and down quarks differ 
in both their mass and charge, 
and it is important to assess the error
thus incurred on $\Delta \alpha$.  
We have studied isospin-breaking effects in $B\to\pi\pi$ decays,
placing particular emphasis 
on the role of $\pi^0-\eta,\eta^\prime$
mixing, as it yields the most significant effects. 
In particular, $\pi^0-\eta,\eta^\prime$ mixing can not only engender 
a $|\Delta I|=5/2$ amplitude, 
breaking the triangle relation of Eq.~(\ref{tri}), 
but also generate a $|\Delta I|=3/2$ penguin amplitude, 
forcing 
$\Delta \alpha_{\rm isospin} - \Delta \alpha \ne 0$.

We recognize that, in nature, all flavor symmetries are
approximate, and we have computed the shift
in $\Delta \alpha$ to leading order in isospin breaking, with an 
assessment of the error in this shift. To realize this, 
we have worked within the QCD factorization framework, though our
results do not depend on the details of such an analysis. 
Rather, the essential point is the utility of 
a combined heavy-quark, $1/m_b$, and 
$\alpha_s$ expansion of the theoretical decay amplitudes. 
We use it to sort through 
the various effects, to determine
that the empirical $B\to\pi\pi$ amplitudes satisfy
a modified triangle relation, Eq.~(\ref{fixedtri}), with a
isospin-breaking parameter $\xi$ which is real, to good approximation. 
Moreover, under the assumption that $\xi$ is real and determined 
exclusively 
by $\pi^0-\eta,\eta^\prime$ mixing, its value can be determined
from experiment, once information on the $\pi^0-\eta,\eta^\prime$
mixing angles is employed. Indeed, the essential improvements
over the analysis of Ref.~\cite{svgpieta} are these: 
that a relationship of form Eq.~(\ref{fixedtri}) exists
with a real parameter $\xi$ and that empirical 
information on $B\to\pi\eta^{(\prime)}$ decays exists and can be
employed to constrain the impact of isospin-breaking effects.
This is important, as the $B\to\eta^{(\prime)}$ form factors
contain contributions which are not constrained by SU(3)$_f$ 
symmetry~\cite{BN}. 
The empirical data on 
$B\to\pi\eta^{(\prime)}$ decays is incomplete, though it 
can be expected to improve. Nevertheless, enough information
currently exists to realize a crucial shift in our perception of
isospin-breaking effects. What matters is not the shift
in $\Delta\alpha$ per se, but 
rather the surety with which we can assess that shift. 
We assess that the change in $\Delta\alpha$ due to isospin-breaking
effects, namely $\delta(\Delta \alpha)\equiv  
\Delta \alpha - \Delta\alpha_0$, where $\Delta\alpha_0$
represents the penguin pollution in the isospin-perfect
limit, is 
\begin{equation}
\delta(\Delta \alpha) = 1.2^\circ\, [\xi] + \,
1.5^\circ \, [P_{ew}]\,
+ \, 1.1^\circ \,[P_{\pi^0-\eta,\eta^\prime}]\, 
+ \dots 
\approx 4^\circ
\,,  
\label{shift}
\end{equation}
where we have resolved the discrete ambiguity in $\Delta \alpha$
by assuming that $\bar\phi > 0$ and $\phi < 0$. We note, currently, that 
$\alpha=(101_{-9}^{+16})^\circ$~\cite{hfag,ckmfitter}; no corrections 
from isospin-breaking effects 
have been included. 
The contribution 
labelled ``$\xi$'' is the shift in $\Delta \alpha$ due to the
presence of a $A_{5/2,2}$ amplitude, realized through 
$\pi^0-\eta,\eta^\prime$ mixing only.
The contributions labelled
``$P_{ew}$'' and ``$P_{\pi^0-\eta,\eta^\prime}$'' represent
the shift in $\Delta \alpha$ due to penguin contributions of effective
$|\Delta I|=3/2$ character. We emphasize that the latter number
is a bound, rather than an explicit estimate. The ellipsis includes
neglected isospin-breaking contributions, such as $A_{5/2,2}$ 
contributions generated by ${\cal O}(\alpha)$ effects on the
$|\Delta I|=1/2$ weak transition, which should be rather 
smaller than the estimate labelled by $\xi$. 
However, the errors in these estimates are smaller 
and are insensitive
to the manner in which the discrete ambiguity in $\Delta \alpha$ 
is resolved: 
\begin{equation}
\sigma_{\alpha}^{\rm IB}= 0.4^\circ\, [\xi] + \,
0.3^\circ\, [P_{ew}] \, 
+ 0.2^\circ \,[P_{\pi^0-\eta,\eta^\prime}]\, + 
1.1^\circ\, [\hbox{bound}] + \dots \approx 2^\circ \,,
\end{equation}
and it is improvable. Note, in particular, the error 
associated with the $A_{5/2,2}$ amplitude comes from doubling the error
in the theoretical computation of $\xi$; here we employ the 
theoretical range in the $F_0^{B\to\eta^{(\prime)}}$ form factors recommended
by Ref.~\cite{BN}. 
This error can be tested, if not mitigated, 
through the use of anticipated empirical data. 
Note, too, that we have included the bound from penguin contributions
to $B^\pm \to \pi^\pm \pi^0$ decays from $\pi^0-\eta,\eta^\prime$ mixing
in our theory error. This can be mitigated with 
improved empirical data. Ultimately, we believe that 
a theoretical systematic error $\sigma_\alpha^{\rm IB}$ 
of ${\cal O}(1^\circ)$
is attainable.

\noindent{\bf Acknowledgments}\,\,
S.G. is supported by
the U.S. Department of Energy under contracts 
DE-FG02-96ER40989 and DE-FG01-00ER45832.


\end{document}